\documentclass[aps,prb,floatfix,superscriptaddress,noshowpacs,twocolumn]{revtex4-2}
\usepackage{subcaption}
\usepackage{graphicx,amsmath,amssymb,epstopdf,wasysym}
\usepackage[space]{grffile}
\usepackage{hyperref}
\usepackage{bm}
\usepackage[dvipsnames]{xcolor}
\usepackage{braket,amsfonts}
\usepackage{dsfont}
\usepackage{float} 
\allowdisplaybreaks

\newcommand{\be}{\begin{equation}}
\newcommand{\ee}{\end{equation}}
\newcommand{\bga}{\begin{gather}}
\newcommand{\ega}{\end{gather}}
\newcommand{\bea}{\begin{eqnarray}}
\newcommand{\eea}{\end{eqnarray}}
\newcommand{\dagga}{{\phantom{\dagger}}}

\newcommand{\bk}{\mathbf{k}}

\newcommand{\br}{\mathbf{r}}
\newcommand{\Ima}{\text{Im}}

\newcommand{\dis}{\displaystyle}

\newcommand{\up}{\uparrow}
\newcommand{\down}{\downarrow}
\newcommand{\fract}[2]{\frac{\dis \;#1\;}{\dis \;#2\;}}

\newcommand{\eqn}[1]{(\ref{#1})}

\newcommand{\ep}{{\epsilon}}

\newcommand{\bw}{\begin{widetext}}
\newcommand{\ew}{\end{widetext}}
\newenvironment{eqs}%
{\begin{equation} \begin{aligned}}%
{\end{aligned} \end{equation} }
\newcommand{\beal}{\begin{eqs}}
\newcommand{\eal}{\end{eqs}}
\newcommand{\bd}[1]{{\boldsymbol{#1}}}
\newcommand{\esp}[1]{\text{e}^{#1}}
\newcommand{\bealn}{\beal\nonumber}

\begin{document}
\title{Revealing spinons by proximity effect}
\author{Antonio Maria Tagliente}
\affiliation{International School for Advanced Studies (SISSA), Via Bonomea 265, I-34136 Trieste, Italy} 

\author{Carlos Mejuto-Zaera}
\affiliation{International School for Advanced Studies (SISSA), Via Bonomea 265, I-34136 Trieste, Italy}
\affiliation{Current: Laboratoire de Physique Th\'{e}orique, CNRS, Universit\'{e} de Toulouse, UPS, 118 Route de Narbonne, F-31062 Toulouse, France} 
\author{Michele Fabrizio}
\affiliation{International School for Advanced Studies (SISSA), Via Bonomea 265, I-34136 Trieste, Italy} 

\begin{abstract} 
The ghost-Gutzwiller variational wavefunction within the Gutzwiller approximation is shown to 
stabilise a genuine paramagnetic Mott insulator in the half-filled single-band Hubbard model.    
This phase hosts quasiparticles that are crucial to the paramagnetic response without showing up in the single-particle spectrum, and, as such, they can be legitimately regarded as an example of Anderson's 
spinons. We demonstrate that these spinons at the interface with a metal 
reacquire charge by proximity effect and thus reemerge in the spectrum as a heavy-fermion band. 
    
\end{abstract}

\maketitle 

\section*{Introduction}
Quantum spin liquids are Mott insulators that host local moments which do not order down to very low temperatures and yet exhibit mutual long-range quantum entanglement well above those temperatures~\cite{Savary_2017,Nocera-Science2020}. The elementary excitations of such systems at energies below the Mott-Hubbard single-particle gap cannot carry electric charge, but they may carry spin as well as possibly emergent quantum numbers~\cite{Savary_2017,Nocera-Science2020,Wen-Topo2013,Sachdev_2019,Sachdev_2023}.    
A celebrated example of these excitations are Anderson's spinons~\cite{PWA-RVB,PWA-PRL1987}, which arise in the
uniform resonating valence bond states proposed to describe highly frustrated 
spin-1/2 Heisenberg models and weakly-doped Mott insulators~\cite{PWA-RVB,PWA-PRL1987,Kivelson-PRB1987,Lee-RMP2006}.
These are neutral gapless spin-1/2 quasiparticles that possess a well-defined \textit{Fermi surface}.
Even though such Fermi surface is likely to suffer from a low-temperature instability opening a finite spin-gap $\Delta_{sp}$~\cite{READ1989609,Read&Sachdev-PRB1990,Hermele-PRB2004,PhysRevB.108.235105,Seifert-NatCom2024}, one could envisage the possibility of a 
spinon degeneracy temperature $T_{sp}$ much larger than $\Delta_{sp}$, 
which would imply that gapless spinons still characterize the physical behaviour for temperatures $\Delta_{sp}\ll T \ll T_{sp}$, or, possibly, at finite magnetic 
fields~\cite{Trivedi-PNAS2019}.   
In those circumstances, a spinon Fermi surface would reveal itself in inelastic 
neutron scattering~\cite{Coldea-PRL2001,Coldea-PRB2003} or thermal 
properties~\cite{Kanoda-NatPhy2008,Yamashita-NatPhys2009,Matsuda-Science2010,Reizo-NatCom2011,Kanigel-PRB2017,Li-PRB2017}. Alternatively, spinons can be detected 
by more local probes, e.g., through their ability to Kondo screen magnetic 
impurities~\cite{Rok-NatPhys2019,Crommie-NatPhys2022,Wang-NatComm2024}, 
or by making proper interfaces with the quantum spin-liquid~\cite{Norman-PRL2009,Berg-PRB2021,Giorgio-NatComm2023,Niklas-PRL2024,Peri-PRB2024}. \\

\noindent
In this work we show that even the simplest interface with a conventional metal 
might be suitable to reveal the spinons hosted in a quantum spin-liquid Mott insulator. Hereafter, we assume a simplified portrait of a quantum spin-liquid, specifically, the paramagnetic Mott insulator that can be stabilized in the single-band Hubbard model forcing spin-$SU(2)$ symmetry within the so-called 
\textit{ghost}-Gutzwiller approximation~\cite{Nicola-ghost}. 
Such interface has already been studied at particle-hole symmetry by Helmes, Costi and Rosch~\cite{Rosh&Costi-PRL2008} using dynamical mean-field theory (DMFT)~\cite{DMFT-review}.  
They showed that the metal penetrates inside the Mott insulator creating a narrow quasiparticle peak that decays exponentially inside the insulating slab over a length that corresponds to the correlation length associated to the Mott transition. This result, later  
reproduced in~\cite{Borghi-PRB2010} using the standard Gutzwiller approximation, was interpreted as a Kondo-like proximity effect: the Mott insulator hosts 
localized moments which are promoted into genuine Kondo resonances by the coupling with the metal~\cite{Rosh&Costi-PRL2008}. While it seems tempting to identify those local moments with spinons \emph{a priori}, such interpretation does not stand up to detailed scrutiny. Indeed, the paramagnetic Mott insulator that is stabilized in 
DMFT has an extensive spin entropy, 
$\ln 2$ per site~\cite{DMFT-review}, as if the local moments were completely free, thus lacking the 
massive entanglement characterizing quantum spin liquids. Exactly the same   
occurs in the standard Gutzwiller approximation~\cite{Brinkman&Rice-PRB1970}, where the Mott insulator 
is a trivial state of independent sites, each occupied by a single electron. 
At the meantime, the DMFT Mott insulator is claimed to exhibit a finite uniform paramagnetic spin susceptibility~\cite{DMFT-review}, which seems not truly consistent with its extensive entropy, nor  
with the evidence that a straight $T=0$ DMFT calculation in the Mott insulator at weak magnetic fields does not converge~\cite{Bauer&Hewson-PRB2007}, barring access to the magnetic susceptibility calculated through the magnetization versus field. The reason of this inconsistency seems to lie in the iterative method through which the DMFT self-consistency equation is solved~\cite{guerci2019}. Indeed,  
to reach convergence in the Mott insulator within the usual iterative scheme, one is forced to assume that the Anderson impurity model onto which the lattice model maps is not described by a pure state, as is expected at $T=0$, but by a mixed one, sum of two states with opposite spin polarizations. \\ 
To address this shortcoming, we decided to use the \textit{ghost}-Gutzwiller approximation~\cite{Nicola-ghost}, which is rigorously 
variational in the limit of infinite lattice-coordination, quite accurate in comparison with DMFT at and away from particle-hole symmetry 
\cite{Nicola-ghost,Carlos-Michele,Nicola-PRB2023,Nicola-PRB2023bis}, and which is able to stabilize a genuine, zero-entropy, paramagnetic Mott insulator~\cite{guerci2019}.
This hints at the existence of spinons,   
whose presence we aim to trace out.
In particular, and to better highlight the chargeless nature of spinons, in this work we     
investigate a metal-Mott insulator interface away from the particle-hole (p-h) symmetric value of the 
chemical potential. For completeness, we start with the standard Gutzwiller approximation (Gut)~\cite{Gutzwiller1963,Gutzwiller1965,Bunemann1998,Fabrizio2007,lanata2015,fluctuations}, 
since the Mott transition changes nature away from p-h symmetry and becomes first order. 
This change of character has been exploited in~\cite{Lee&Yee-PRB2017} to 
study by DMFT the interface between metal and Mott insulator phases coexisting at equilibrium. 
We next move to the main part of our work, namely, the interface studied by the \textit{ghost}-Gutzwiller approximation (gGut)~\cite{Nicola-ghost,guerci2019,frank2021,Lanata2022,Mejuto2023a,Guerci2023,Lee2023a,Lee2023b,Mejuto2024,Lee2024}. Specifically, the paper is organised as follows: in section~\ref{method} we give a brief introduction
to the Gutzwiller and \textit{ghost}-Gutzwiller approximations for the Metal-Mott interface. In section \ref{singleghost}, we 
present results using the standard Gut and show how the phenomenology
of the Kondo proximity effect extends to non particle-hole symmetric cases.
In section~\ref{pinning} we study the same model by the gGut.
Finally, in section~\ref{bulk}, we discuss the results by solving analytically the 
gGut variational problem for a bulk Mott insulator in the limit of a very large Mott-Hubbard gap. 
In that limit, we also show the solution forcing a finite spin-polarization, which clearly 
unveils the role of spinon in allowing the Mott insulator to sustain partial spin-polarization, and 
gives access to the paramagnetic susceptibility. Section~\ref{Conclusions} is devoted to 
conclusions.\\  
\section{Model and methods}
\label{method}
In Fig.~\ref{systemfig} we show the system under investigation,
 which consists of a set of metallic layers, represented as weakly interacting Hubbard models,
 coupled to several Mott insulating ones, strongly interacting Hubbard models. The system has a total of $N$ layers and the interlayer coupling is just a single particle hopping.
 We assume that intralayer and interlayer hopping strengths are equal, while the difference between metallic and Mott-insulating layers is modelled by a layer-dependent Hubbard repulsion.
The system is therefore described by the Hamiltonian:
\begin{figure}[ht]
    \centering
    \includegraphics[width=0.5\textwidth]{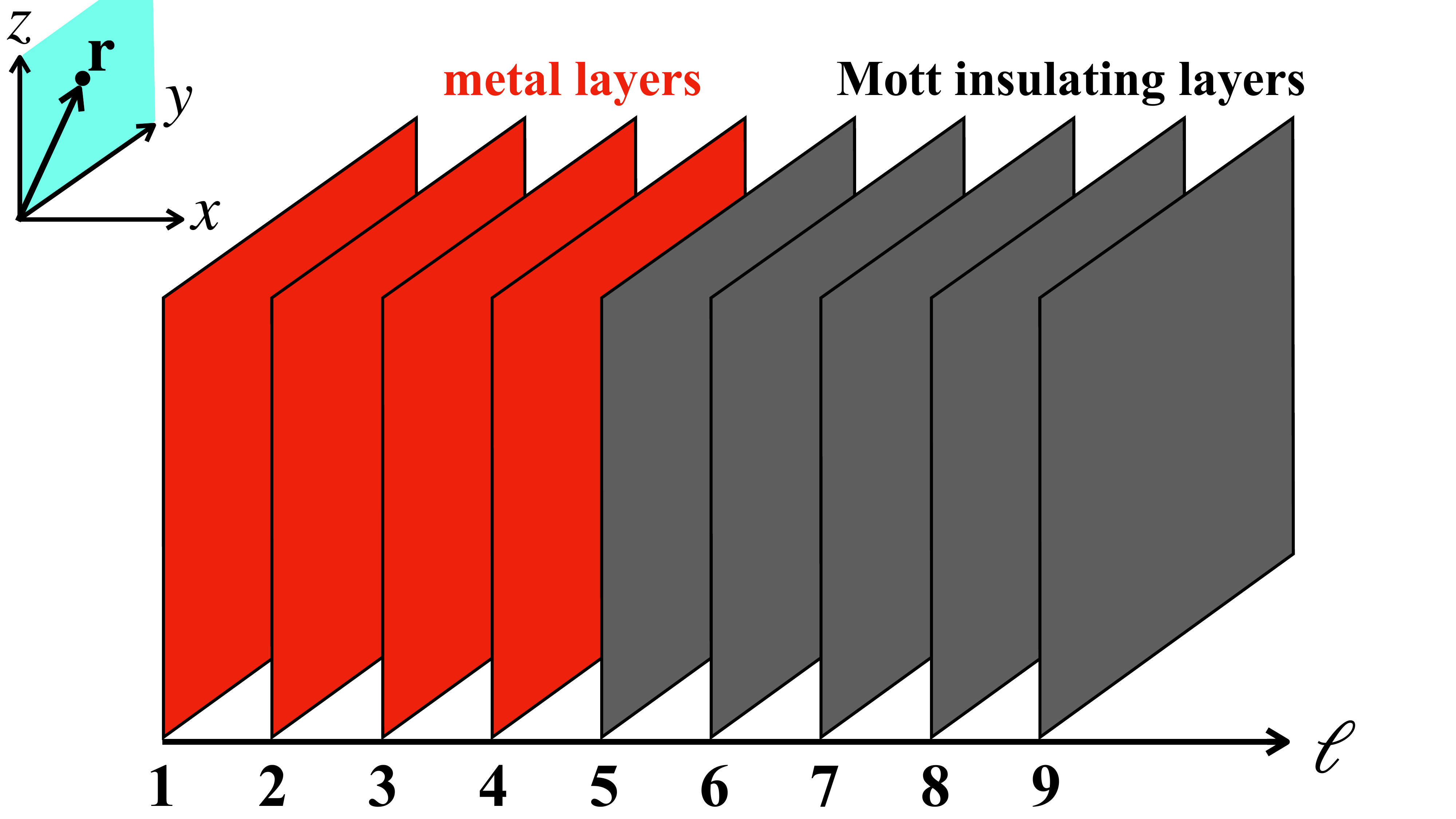}
    \caption{Schematic representation of the slab geometry in the case of 
    9 layers, 4 of which are metallic, in red, and the rest Mott insulating, in grey. Layers are stacked along the $x$-axis, while a site in each layer is identified by the vector $\br$ in the $y$-$z$ plane.
     }\label{systemfig}
\end{figure}

\beal
\mathcal{H} &= \sum_{\ell=1}^{N}\, \sum_{\bk\sigma}\,\epsilon_{\bk}\,
     c^{\dagger}_{\ell\bk\sigma}\,c^\dagga_{\ell\bk\sigma}\\
&\quad    -t\,\sum_{\ell=1}^{N-1}\, \sum_{\bk\sigma}\,\big( 
c^{\dagger}_{\ell+1\bk\sigma}\,
c^\dagga_{\ell\bk\sigma} + H.c.\big) \\
&\qquad + \sum_{\ell=1}^{N}\,\sum_{\br} \bigg\{\frac{U_\ell}{2}\,\big(n_{\ell\br} -1\big)^2 -\delta \mu\, n_{\ell\br}\bigg\}\,,\label{Ham}
\eal    

where the lattice sites are identified by an intralayer coordinate $\br$ and a layer index $\ell=1,\dots, N$, and thus $n_{\ell\br}$ is the on-site occupation number, while the non-interacting intra-layer dispersion  
$\epsilon_{\bk} = -2t\,(\cos k_y + \cos k_z)$, with $\bk$ the momentum in the 
$y$-$z$ plane, $\sigma$ labels the spin of the electrons. Hereafter, we take $t=1$ as energy unit. We remark that $\delta\mu$ in Eq.~\eqn{Ham} parametrizes the deviation of each layer with respect to the p-h symmetric case, $\delta\mu=0$, while the physical layer-dependent \textit{chemical} potential is in reality $\mu_\ell=U_\ell/2+\delta\mu$. \\
As mentioned in the introduction, we study the system by means of the Gutzwiller (both standard and ghost) approximation~\cite{Borghi-PRB2010,Nicola-ghost}.
The Gutzwiller variational wavefunction $\ket{\Psi}$ consists of an uncorrelated variational Slater determinant $\ket{\Psi_*}$, defined in a (possibly enlarged) auxiliary Hilbert space, which is projected by 
a linear operator $\mathcal{P}_G$ onto the physical Hilbert space. 
The conventional Gutzwiller wavefunction~\cite{Gutzwiller1963,Gutzwiller1965} corresponds to the case in which the dimensions of the physical and auxiliary Hilbert spaces coincide, while, if the auxiliary Hilbert space is larger than the physical one, we talk about the \textit{ghost}-Gutzwiller wavefunction~\cite{Nicola-ghost}. The inclusion of additional auxiliary fermions extends the variational freedom and gives access to high-energy spectral features such as the Hubbard bands~\cite{Nicola-ghost}.\\
For the case of the Hamiltonian in Eq.~\eqn{Ham} the variational wavefunction reads:
\beal
\ket{\Psi} &= \mathcal{P}_{G}\ket{\Psi_*} \,,\\
\mathcal{P}_{G} &= \prod_{\br,\ell} 
   \Bigg\{\sum_{\gamma,\Gamma}\, \lambda_{\gamma,\Gamma}(\ell) \ket{\br,\ell;\gamma}
   \bra{\br,\ell;\Gamma}\Bigg\} \,.\label{gutansatz}
\eal
The states $\Gamma$ are local Fock states of the auxiliary fermions, whose annihilation operators we denote as $d^\dagga_{\ell\br\alpha\sigma}$, where $\alpha=1,\dots,N_{aux}$ span the auxiliary spinful orbitals. On the contrary, the set $\gamma$ includes the local physical Fock states.\\ 
In Eq.~\eqn{gutansatz} we have assumed translational invariance within each layer, 
so that our variational parameters depend only on the $\ell$ coordinate. This also 
implies that the variational wavefunction Eq.~\eqn{gutansatz} cannot develop the antiferromagnetic order parameter that characterizes the actual 
ground state of Eq.~\eqn{Ham} at finite on-site repulsion. Such restriction into a subspace of the whole 
Hilbert space, which does not include the expected ground state in the specific example, 
is easy to implement in the variational approach.
\\
Following~\cite{lanata2015}, we can associate the parameters $\lambda_{\gamma,\Gamma}(\br,\ell)$ to the components of a wavefunction that describes an impurity coupled to $N_{aux}$ baths. In the slab geometry, we need an impurity 
wavefunction $\ket{\phi(\ell)}$ for each layer. 
Using the Gutzwiller approximation, specified in Appendix~\ref{gutappendix}, we can write the variational energy as:
\be
    E[\Psi] = \bra{\Psi_*}\mathcal{H}_*\ket{\Psi_*} + \mathcal{A}\sum_\ell\, \bra{\phi(\ell)} \mathcal{H}_{loc}(\ell)\ket{\phi(\ell)}\,,\label{gutenergy}
\ee
with $\mathcal{A}$ the number of sites within each layer, 
\be
\mathcal{H}_* = \sum_{\bk \ell \ell'\sigma}\,\sum_{\alpha \beta}\,
d^\dagger_{\ell\bk\alpha\sigma}\;
   R_{\alpha\sigma}^{\dagger}(\ell)\,t_{\ell\ell'}(\bk)\,
   R_{\beta\sigma}(\ell')\;
   d^\dagga_{\ell' \bk \beta \sigma}\,,\label{hqp}
\ee
where $t_{\ell\ell'}(\bk) = \ep(\bk)\,\delta_{\ell\ell'} 
-t\,\sum_{p=\pm 1}\,\delta_{\ell',\ell+p}$, 
and
\beal   
\mathcal{H}_{loc}(\ell) = \frac{U_\ell}{2}\big(n_\ell -1\big)^2 
-\delta \mu\, n_\ell\,,\label{hloc}
\eal
with $n_\ell$ the occupation number of the impurity level. 
The vector $\bd{R}(\ell)$ with components $R_{\alpha\sigma}(\ell)$ in Eq.~\eqref{hqp},
defined in Eq.~\eqref{R} of the Appendix~\ref{gutappendix}, see also~\cite{fluctuations}, looks like a wavefunction renormalisation, 
\bealn
c^\dagga_{\ell\bk\sigma} \to \sum_\alpha\,
R_{\alpha\sigma}(\ell)\,d^\dagga_{\ell\bk\alpha\sigma}\,,
\eal
suggesting that $d^\dagga_{\ell\bk\alpha\sigma}$ can be identified with the quasiparticle Fermi operators. \\
Using the impurity formulation, we can minimize self-consistently~\cite{Nicola-ghost} the energy in Eq.~\eqref{gutenergy}.
In particular, 
the Slater determinant and the impurity wavefunctions are respectively the ground states of a quasiparticle Hamiltonian $\mathcal{H}^{qp}$, and of the impurity Hamiltonians
$\mathcal{H}_{imp}(\ell)$, with expressions:
\beal
\mathcal{H}^{qp} &= \mathcal{H}_* -\sum_{\bk\ell\sigma}\,\sum_{\alpha\beta}
d^\dagger_{\ell\bk\alpha\sigma}\,\lambda_{\alpha\sigma,\beta\sigma}(\ell)\,
d^\dagga_{\ell\bk\beta\sigma}\\
&= \sum_{\bk\sigma}\, \mathcal{H}^{qp}_{\bk\sigma}\,,\label{qpham}
\eal
and
\beal
\mathcal{H}_{imp}(\ell) &= \mathcal{H}_{loc}(\ell) 
+ \sum_{\alpha\sigma}\,V_{\alpha\sigma}\,\big(d^\dagger_{\ell\alpha\sigma}\,
c^\dagga_{\ell\sigma}+H.c.\big)\\
&\qquad -\sum_{\alpha\beta\sigma}\,
d^\dagger_{\ell\alpha\sigma}\,\lambda^c_{\alpha\sigma,\beta\sigma}(\ell)\,
d^\dagga_{\ell\beta\sigma}\,.\label{hamimp}
\eal
The self-consistency condition is expressed in terms of the one-body reduced density matrices of the impurity model and of the quasi-particle Hamiltonian:
\be
\bra{\phi(\ell)} d^\dagga_{\ell\alpha\sigma}\,d^\dagger_{\ell\beta\sigma}  \ket{\phi(\ell)} = \bra{\Psi_*} d^\dagger_{\ell\br\beta\sigma}\,
 d^\dagga_{\ell\br\alpha\sigma} \ket{\Psi_*}\,.
 \label{Gut-self-consistency-eq}
\ee
The Hamiltonians Eq.~\eqref{qpham} and Eq.~\eqref{hamimp} depend on a set of Lagrange multipliers
$\lambda^c_{\alpha\sigma,\beta\sigma}(\ell)$, $\lambda_{\alpha\sigma,\beta\sigma}(\ell)$ and $V_{\alpha\sigma}(\ell)$, whose expressions are given in  Appendix~\ref{gutappendix}; more details about the derivation and implementation
of the Gutzwiller equations can be found, e.g., in~\cite{fluctuations,Nicola-ghost,Mejuto2023a}.
The parameter $V_{\alpha\sigma}(\ell)$ is the hybridisation between the impurity and the baths, while $\lambda^c_{\alpha\sigma,\beta\sigma}(\ell)$ and $\lambda_{\alpha\sigma,\beta\sigma}(\ell)$ are one-body 
potentials for the bath levels and the quasiparticles, respectively. 
It is worth emphasising that the physical chemical potential 
$\mu=U/2+\delta\mu$ translates into the energy level of the impurity, see Eq.~\eqn{hloc}. Through the energy minimization and the self-consistency equation~\eqn{Gut-self-consistency-eq}, $\mu$ plays a role in determining the local potential $\lambda_{\alpha\sigma,\beta\sigma}(\ell)$ of the quasiparticle Hamiltonian in Eq.~\eqref{hqp}, and thus the position of the 
quasiparticle bands. \\
Through the knowledge of the quasiparticle Hamiltonian, we can derive an approximate expression for the Green's function matrix for each layer, 
\be
    G(\omega,\ell,\bk) = \bm{R}^{\dagger}(\ell) \,\Bigg(\frac{1}{\omega + i0^+ -\mathcal{H}^{qp}_{\bk\sigma}}\Bigg)_{\ell\ell}\bm{R}(\ell)\,,
    \label{greenfunction}
\ee
which allows defining an approximate self-energy $\Sigma(\omega,\ell,\bk)$ and a layer-dependent quasiparticle residue:
\begin{align}
    Z(\ell,\bk) = \left(1 - \frac{\partial \Sigma(\omega,\ell,\bk)}{\partial \omega}{\bigg|_{\omega = 0}}\right)^{-1}~.
    \label{eq:Z}
\end{align}
Other local observables can be obtained through the impurity wavefunction. In particular, we will be interested in the electron density per layer given by:
\bealn
    n_{\ell} = \sum_\sigma\,
    \bra{\phi(\ell)} c^\dagger_{\ell\sigma}\,
    c^\dagga_{\ell\sigma}\ket{\phi(\ell)}\,.
\eal
\section{Wetting critical behaviour in the standard Gutzwiller approximation}\label{singleghost}
\begin{figure}[h!]
    \centering
    \includegraphics[width=0.9\linewidth]{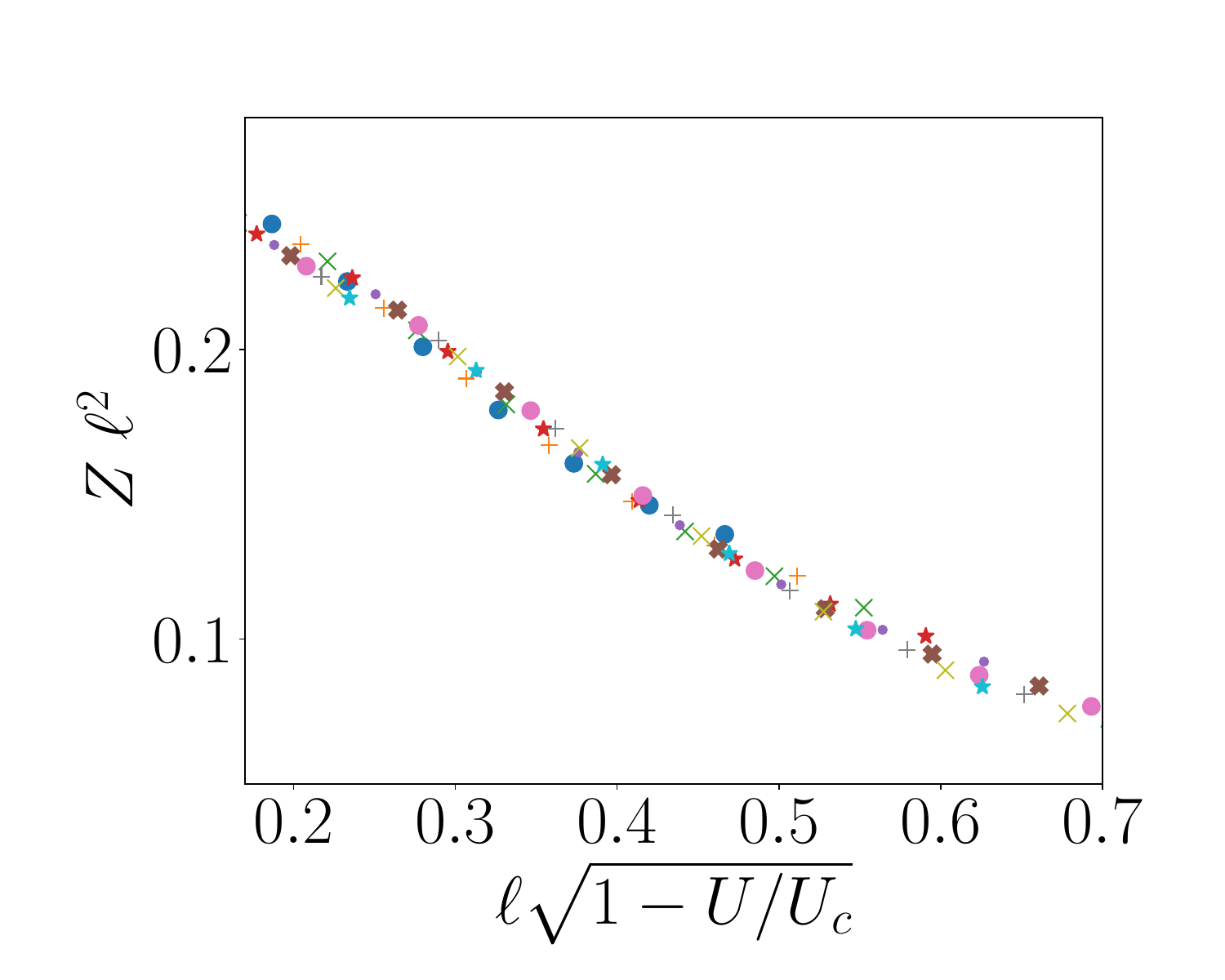} 
    \caption{Scaling plot for the quasiparticle residue at $\delta\mu=-0.9$ close to the critical point confirming the mean field exponent $\nu = 1/2$.}
    \label{fss}
\end{figure}

Using the variational ansatz in Eq.~\eqref{gutansatz} with $N_{aux} = 1$, thus the standard Gutzwiller wavefunction, we can study the fate of the Kondo-like proximity effect away from \mbox{p-h} symmetry. 
In the standard Gut, just as in DMFT, the Mott transition away from particle-hole symmetry becomes genuinely first-order. In other words, there is a range 
of $U\in [U_{c1},U_{c2}]$, bounded by the insulator and metal spinodal points, 
$U_{c1}$ and $U_{c2}>U_{c1}$, respectively, where the two phases coexist. Their corresponding energies cross at $U_c$ in between the spinodal values. At particle-hole symmetry, $U_c$ accidentally coincides with $U_{c2}$ in DMFT, while, in standard Gut, the transition turns second order.\\  
One may wonder how such change of character in the transition affects the proximity effect 
observed in~\cite{Rosh&Costi-PRL2008,Borghi-PRB2010}. It is known that approaching a first-order phase transition one can still find a surface critical phenomenon, the wetting transition~\cite{wetting}.
It was conjectured in~\cite{wetting-ising}, using an effective spin model for the Mott transition, that the Kondo proximity effect becomes an example of total wetting~\cite{degennes} when the Mott transition is discontinuous.
\begin{figure}[h!]
    \centering
    \includegraphics[width=0.9\linewidth]{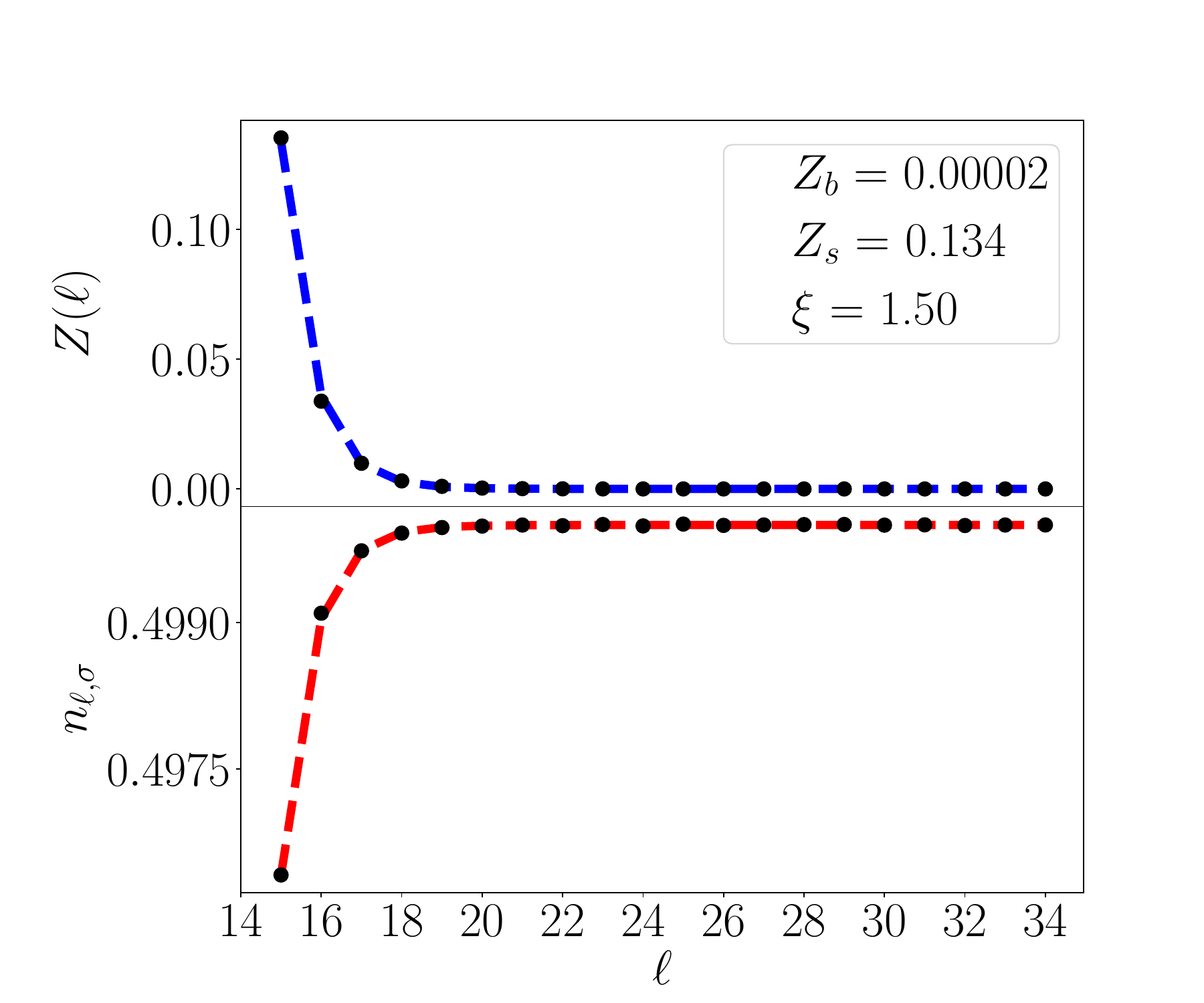}
    \caption{Spatial profile of the quasiparticle residue and electron density per spin across the layers.
     The last metallic layer is placed at $\ell = 14$. The Hubbard repulsion in the metal layers 
is $U=2$  and in the insulating ones $U = 16.828$, and $\delta\mu = -0.9$. The dashed line in the first panel is obtained by the fit, instead 
in the dashed line in the second panel is obtained through \ref{delta_x}. The legend shows the parameter obtained by the fit, where $Z_s = R_s^2$ and $Z_b = R_b^2$
are the value of the metallicity in the first insulating layer and deep in the bulk.}\label{z-n}
\end{figure}
We quantitatively address this question by simulating a metal-Mott interface away from p-h symmetry. More specifically, we study a system of 100 layers where 15 layers are metallic with 
$U = 2$, the value that we use here and in the next section. In Fig.~\ref{z-n} we plot the layer-dependent quasiparticle residue and the electron density per spin 
starting from the last metallic layer and for $\delta\mu = -0.9$, see Eq.~\eqn{Ham}.
Differently from the \mbox{p-h} symmetric case, the proximity effect is shown also in the electron density, which is not anymore constrained by p-h symmetry.
We fit the layer dependent wavefunction renormalisation through the expression:
\beal
R(\ell) &= R_{b} + \big(R_{s} -R_{b}\big)\,\esp{(\ell-\ell_*)/\xi}\;,\\
\label{fitting formulas}
\eal
where $\ell_*$ is the index of the first insulating layer and $\xi$ the correlation length.
From equation \ref{fitting formulas} we can obtain the quasiparticle residue, which in standard Gutzwiller is just $Z(\ell) = R(\ell)^2$.
Instead, the functional form of the density can be deduced, near criticality, by \ref{fitting formulas} through equation \ref{delta_x} in appendix \ref{continuum}.
For $\delta\mu = -0.9$, the first-order Mott transition occurs at ${U_c \approx 16.3}$, and $\xi$ diverges as $1/\sqrt{|U-U_c|}$ approaching 
the transition, a typical mean-field exponent. To further confirm the critical properties of the wetting layer, we rescale both the layer coordinate and $Z(\ell)$ 
as discussed in~\cite{Rosh&Costi-PRL2008}. Indeed, all data for different values of $U$ collapse on the same curve, as shown in Fig.~\ref{fss}, supporting the wetting critical behaviour, see also  Appendix~\ref{continuum}.
\section{Ghost Gutzwiller results}\label{pinning}
\begin{figure}[h!]
    \centering
    \includegraphics[width=0.5\textwidth]{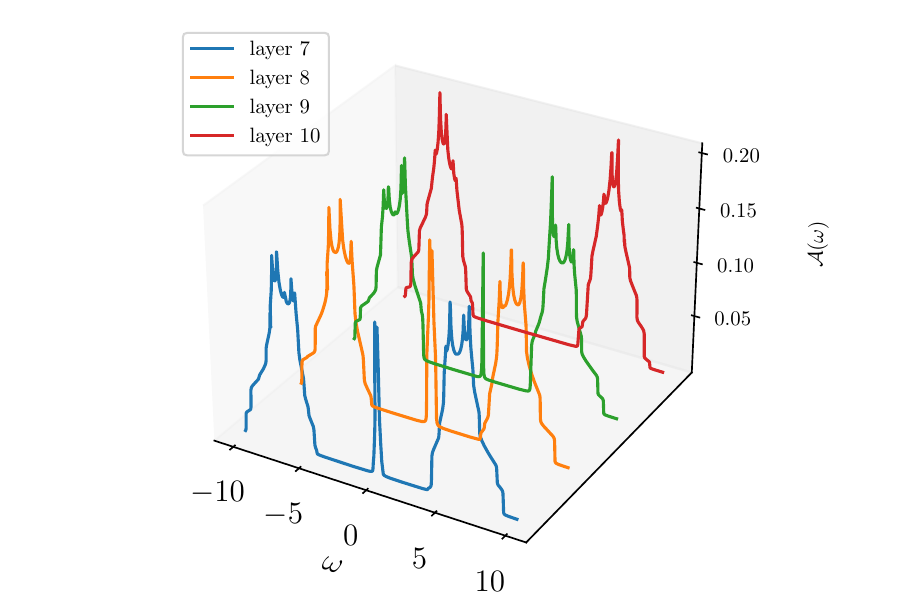}
    \caption{Layer resolved spectral function for $\delta\mu=0.5$ and $U=14.316$ showing how the Kondo peak, always centred at zero frequency, disappears moving away from the interface.}
    \label{layerspectral}
\end{figure}

We can readily extend the results of the previous section to the gGut approximation. In the bulk case and for the single band Hubbard model, it has been shown~\cite{Nicola-ghost} that 
already with $N_{aux} = 3$ one can obtain the Hubbard bands and have a quantitative description of the Mott transition in excellent agreement with the DMFT solution.\\
We find that moving away from p-h symmetry the transition within gGut remains second-order, even upon increasing the number of
auxiliary fermions, as discussed in section~\ref{bulk}. We are therefore back to 
the surface critical behaviour discussed in~\cite{Rosh&Costi-PRL2008,Borghi-PRB2010}. 
\begin{figure}[h!]
    \centering
    \includegraphics[width=0.9\linewidth]{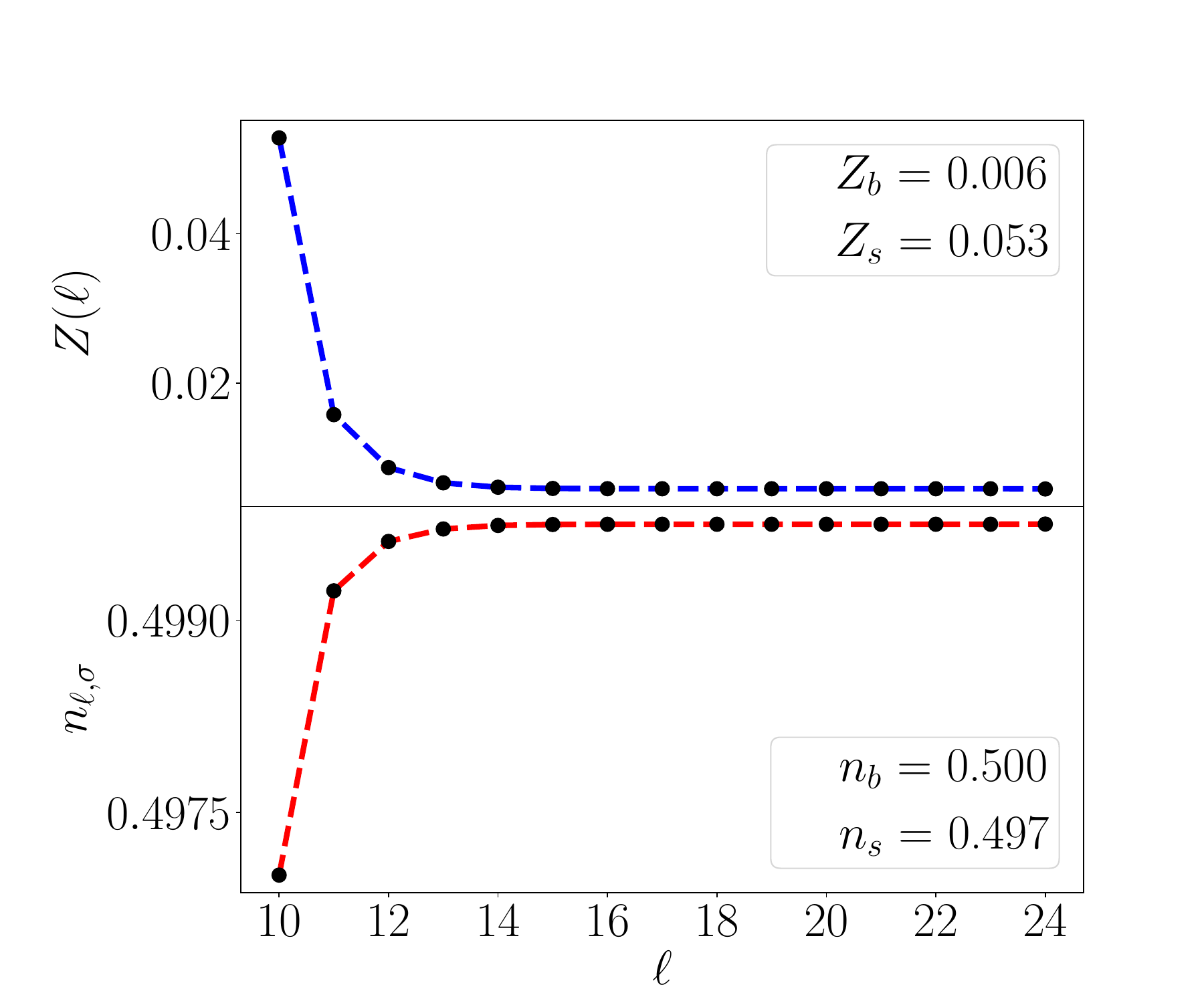}
    \caption{Behaviour of the gGut quasiparticle residue and electron density per spin at $\delta\mu = -0.5$, using  
    $U = 14.385$ in the insulating layers. The last metallic layer is $\ell=9$. In the legends are reported the quasiparticle residue and the density in the bulk ($Z_b$, $n_b$) and in the first insulating layer ($Z_s$,$n_s$).}\label{zn-ghost}
\end{figure}
We study systems up to 30 layers with $N_{aux} = 3$ and find a clear proximity effect both in the quasiparticle
residue and the electron density, as shown in Fig.~\ref{zn-ghost}.
\begin{figure}[h!]
    \centering
    \includegraphics[width=0.9\linewidth]{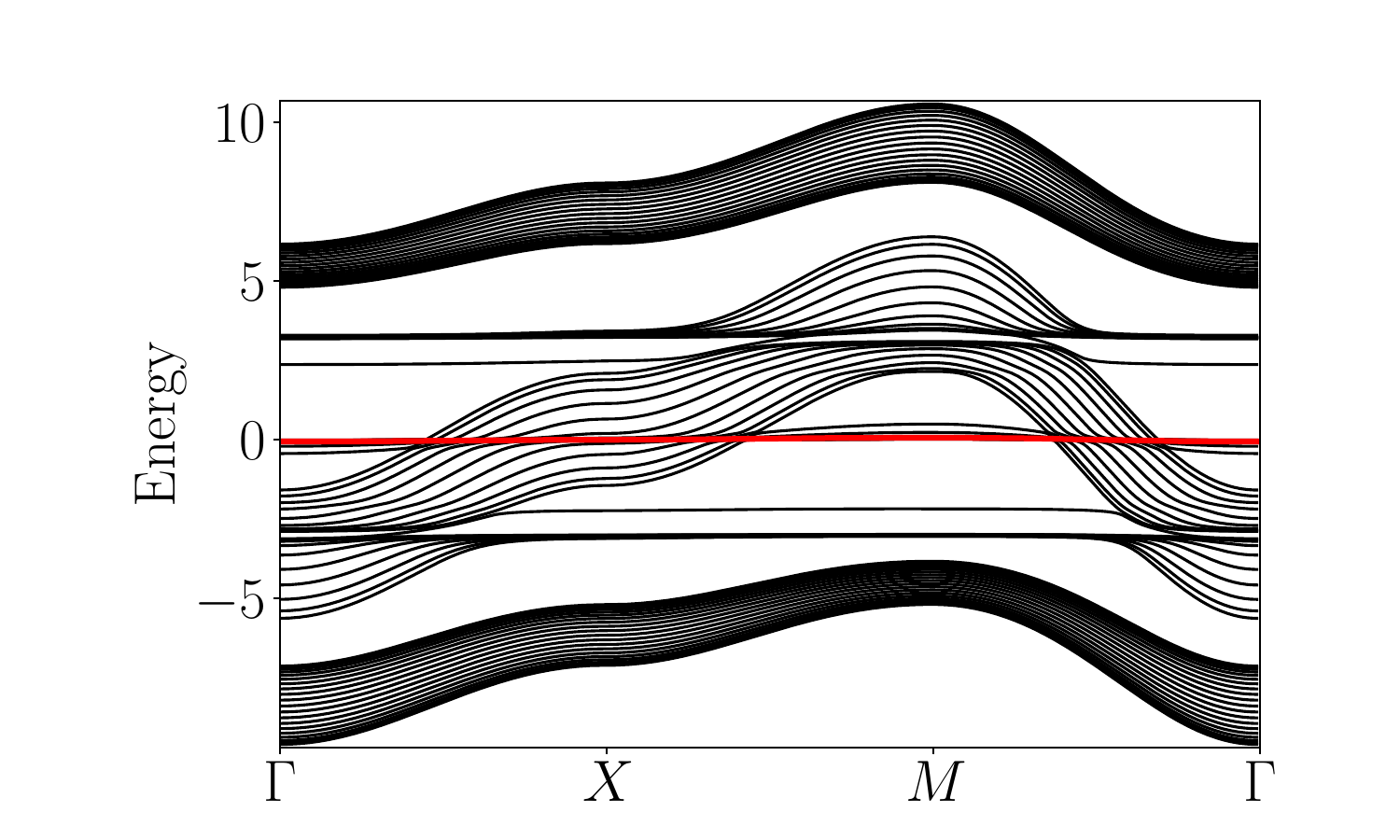}
    \caption{Bands of the quasiparticle Hamiltonian for a slab with 10 metallic layers and 20 Mott insulating ones, at $\delta\mu= -0.5$ and $U=15$ in the insulating layers.}\label{zeros}
\end{figure}
Besides uncovering, as before, the critical behaviour, we can now examine how the presence of the metallic leads affects the quasiparticles in the insulating layers. In Fig.~\ref{zeros}, we show the quasiparticle bands along high-symmetry 
paths of the two-dimensional Brillouin zone, corresponding to the $y$-$z$ plane, in 
the case with 10 metallic layers out of 30. 
We observe the following qualitative structure of the figure: 
some bands cross zero energy, corresponding to the metallic layers, and the others accumulate to form lower and upper Hubbard bands. In particular, a feature that we always find is  
a set of flat bands pinned at zero energy. We shall thoroughly discuss these zero-energy modes in the Sec.~\ref{bulk}, but we anticipate here that they essentially describe spinons, i.e., gapless spin-1/2 neutral excitations. What is remarkable in Fig.~\ref{zeros} 
is that the spinon band does not shift rigidly with $\delta\mu$ in Eq.~\eqn{Ham}, as the Hubbard bands instead do, but remains pinned at zero energy, consistently with the spinons being neutral quasiparticles and thus unaffected by the chemical potential. \\
The spinons thus offer a very simple interpretation 
of the Kondo proximity effect: at the interface with the metal the spinon reacquires charge and gets promoted into a Kondo resonance narrowly peaked at zero 
energy. More precisely, the flat band hybridizes with the last metallic layer 
like in heavy fermion systems, leading to a very narrow band and the opening of small 
hybrization gaps. To motivate this claim we plot in figure \ref{resbands} the layer character of the low energy bands of figure \ref{zeros}, showing how the first insulating layer
contribute to the dispersive bands.
\begin{figure}[h!]
    \centering
    \includegraphics[width=0.5\textwidth]{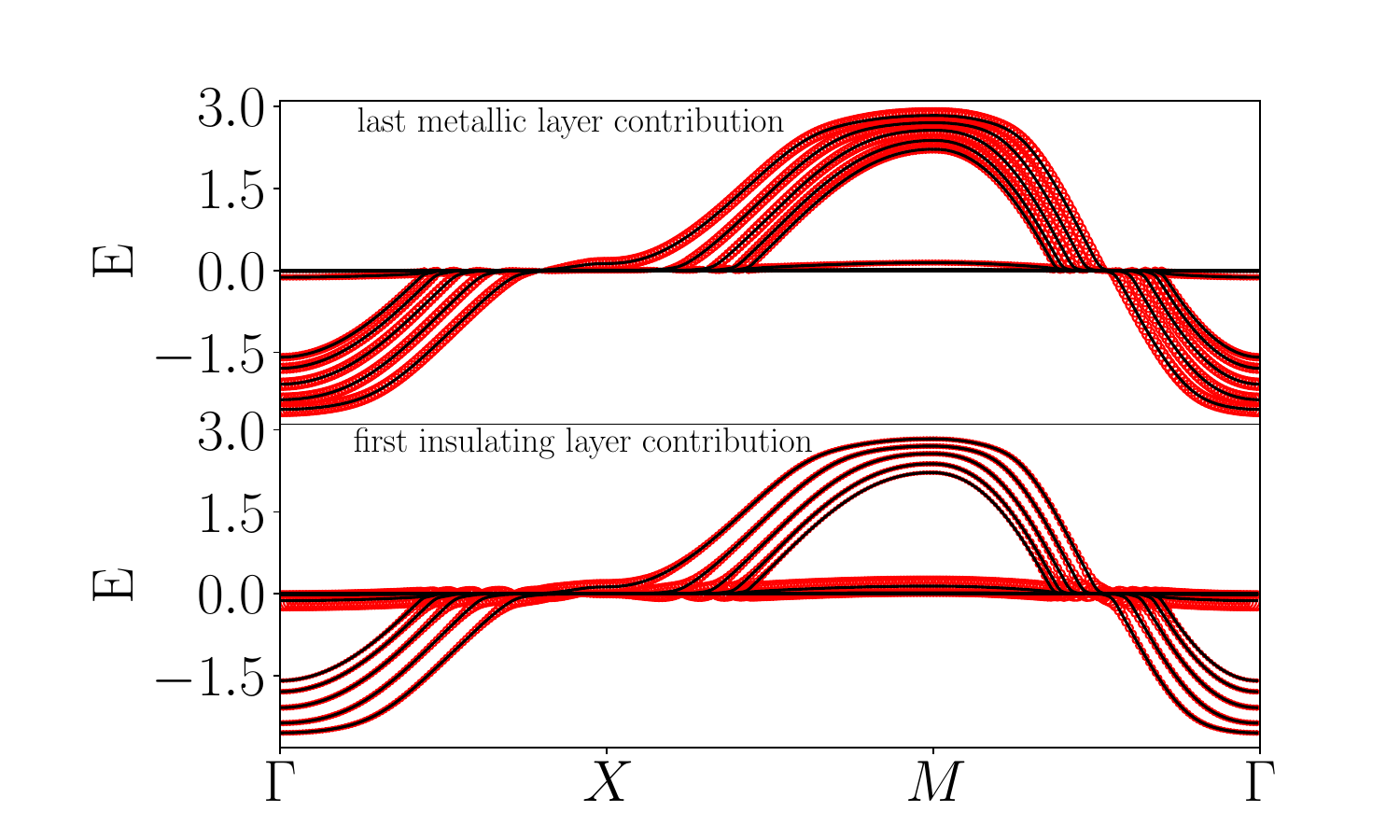}
    %\vspace{-1cm}
    \caption{The layer-resolved bands of figure \ref{zeros} at low energy, where the size of the red circles is proportional to the contribution of the corresponding layer. Top panel: contribution of the last metallic layer.
    Bottom panel: contribution of the first insulating layer. We observe that the flat band of the insulating layer acquires a finite dispersion through the hybridization with the neighbouring metallic one.}
    \label{resbands}
\end{figure}
For completeness, in Fig.~\ref{layerspectral} we plot the layer-resolved 
spectral function $\mathcal{A}(\omega,\ell) = -\Ima\,G(\omega,\ell)/\pi$ 
for a slab made of five metallic and five insulating layers that further highlights the layer evolution of the Kondo peak. 
We have thus shown that the zero-energy quasiparticle excitations inside the paramagnetic Mott insulator, a simple version of quantum spin liquid, appear at metallic interfaces in the form of a heavy-fermion proximity effect.

\section{Origin of the spinon mode}\label{bulk}
\begin{figure}[h!]
    \centering
    \includegraphics[width=0.9\linewidth]{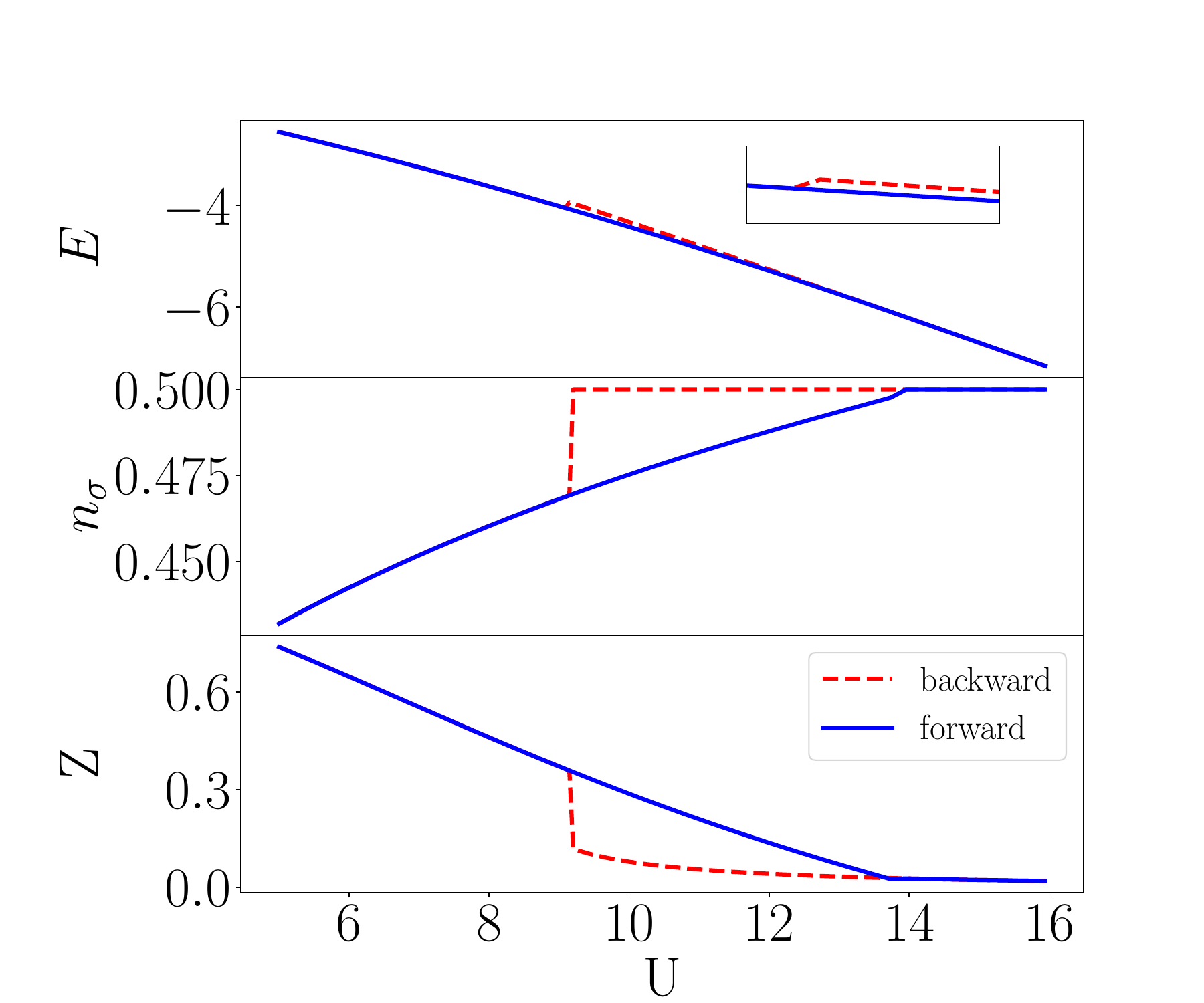}
    \caption{Ground state energy, quasiparticle residue, and density per spin as a function of $U$ in the cubic lattice Hubbard model at $\delta\mu=-1$.}\label{coexistence}
\end{figure}

\begin{figure}[h!]
    \centering
    \includegraphics[width=0.9\linewidth]{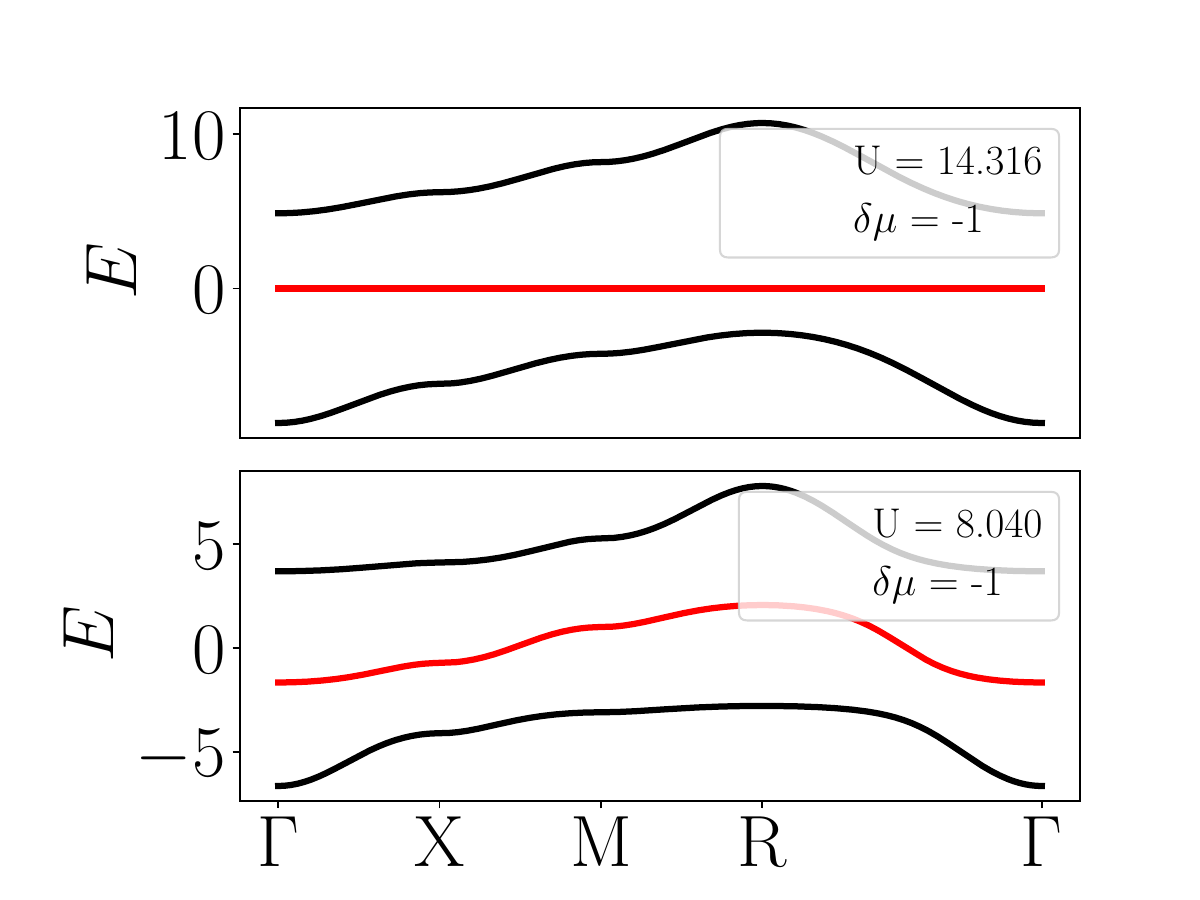}
    \caption{Bands of the quasiparticle Hamiltonian drawn along high-symmetry paths in the Brillouin zone of a cubic lattice for $\delta\mu = -1$. The top panel refers to $U = 14.316$ in the Mott insulating phase, while the bottom panel to $U = 8.040$ in the correlated metal with preformed Hubbard bands.}\label{bands}
\end{figure}

In this section, we highlight by simple analytical calculations the role played by the 
zero-energy quasiparticles in the variational optimization, and show that they possess all prerequisites for being legitimately regarded as \textit{spinons}. Consistently with the slab geometry 
in Fig.~\ref{systemfig}, we consider a bulk system on a cubic lattice with periodic boundary conditions. \\
We start by showing that in gGut the Mott transition is  
second order also away from p-h symmetry, unlike in Gut or DMFT. In Fig.~\ref{coexistence} we show the evolution of the energy, average density per spin and quasiparticle residue, first upon increasing $U$ from the weakly correlated metal (blue solid lines), and then decreasing it from the insulating side (red dashed lines). We observe a metal-insulator coexistence, but, differently from DMFT away from p-h symmetry, the energy does not show any discontinuity in the slope as expected at a first order phase transition. Rather, the behaviour in Fig.~\ref{coexistence} resembles the DMFT one at p-h symmetry~\cite{DMFT-review}, where there is 
metal-insulator coexistence but the metal spinodal point coincides with the value of $U$ at which the 
two energies cross, hence the accidental second-order transition. We mention that a similar disagreement 
between gGut and DMFT was observed in~\cite{guerci2019} studying the Mott transition in presence of 
a magnetic field, which is predicted continuous in gGut and discontinuous in DMFT. 
Another feature of the bulk gGut solution, in common with the previous slab geometry, is that the quasiparticle bands include dispersing and p-h asymmetric lower and upper Hubbard bands, 
as well as a flat band pinned at zero energy in the Mott phase that reacquires a narrow dispersion in the correlated metal, see Fig.~\ref{bands}. 
\\

\subsection{Paramagnetic solution}
\begin{figure}[h!]
    \centering
    \includegraphics[width=0.45\textwidth]{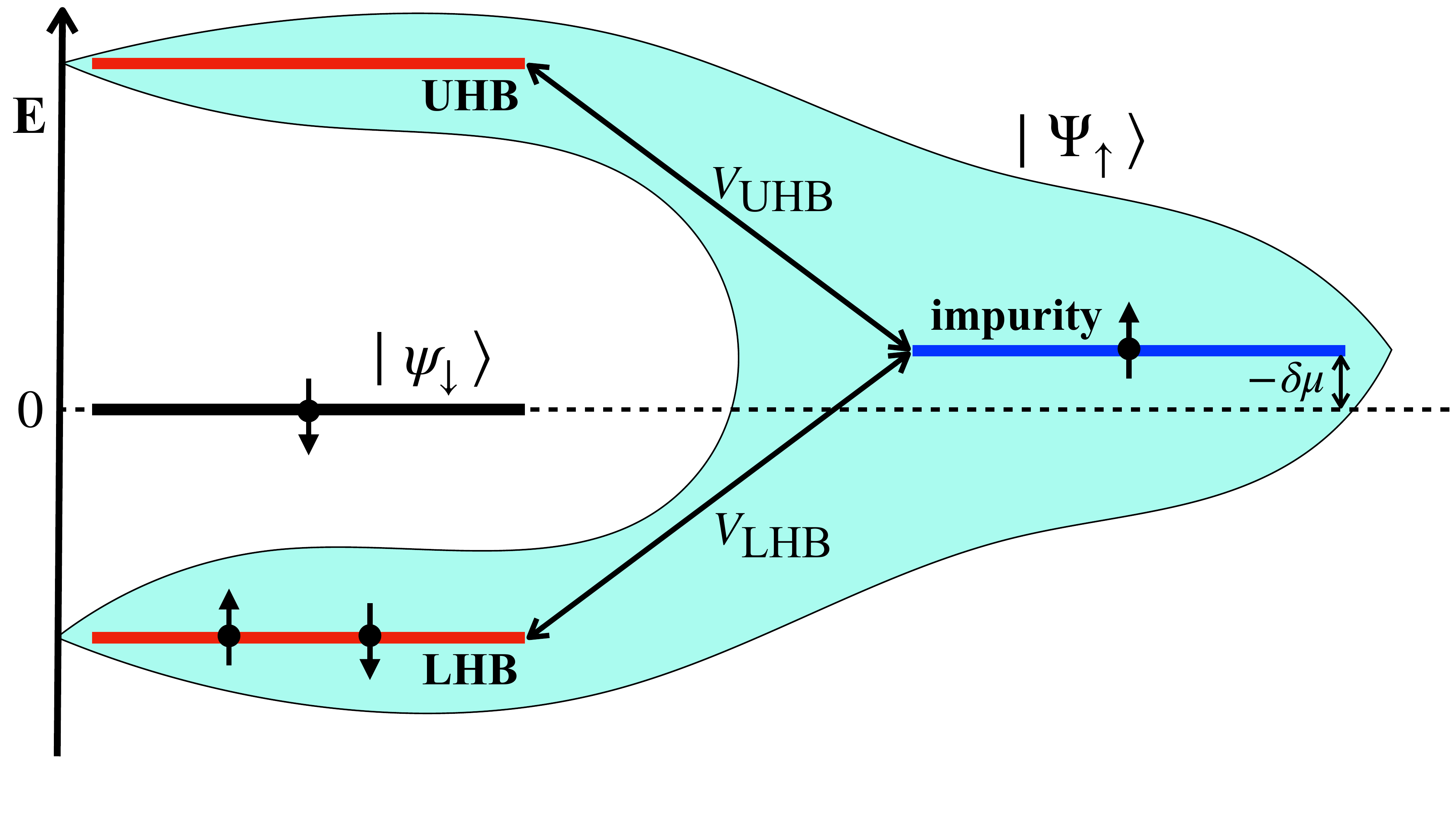}
    \caption{Schematic representation of the effective impurity model in the Mott insulator.}
    \label{scheme}
\end{figure}

To enlighten the role of the zero-energy quasiparticles, 
we here elaborate on the arguments presented in~\cite{guerci2019}, 
see also~\cite{Daniele-Thesis}. Specifically, we consider the embedded impurity model with 
$N_{aux}=3$, sketched in Fig.~\ref{scheme}. This is representative of all models with $N_{aux}$ odd, while the case of even $N_{aux}$ will be discussed later. The impurity level is on the right, in blue, shifted up in 
energy by $-\delta\mu$, with $\delta\mu<0$ in the figure. In the Mott insulator, $U$ overwhelms by far $\delta\mu$ so that the 
impurity wants to be half-filled, in the figure with a spin-up electron. 
The positive-energy and negative-energy bath levels, in red, correspond, respectively, to the 
lower (LHB) and upper (UHB) Hubbard bands, and are hybridised with the impurity by 
$V_\text{LHB}$ and $V_\text{UHB}$. The bath level in between, in black, is instead not hybridised with the 
impurity~\cite{Nicola-ghost}. At order zero in $V_\text{LHB}$ and $V_\text{UHB}$, the LHB is full and the 
UHB empty, as in the figure. However, because of the finite hybridisation, the actual ground state 
of the impurity+LHB+UHB model, in cyan in Fig.~\ref{scheme}, includes other configurations, for instance the impurity empty and 
one electron on the UHB. We remark that, to maximise the energy gain of the hybridisation, it is preferable that the LHB and UHB are symmetrically located with respect to the impurity level, thus both 
shifted by $-\delta\mu$, as in the figure.\\
We denote as $\ket{\Psi_\sigma}$ the actual ground state of the impurity+LHB+UHB model, where $\sigma$, $\up$ in the figure, is the $z$-component of the total spin, which is 1/2 in the ground state. One may na\"{\i}vely argue that the decoupled bath level is fully irrelevant and can be in any configuration, singly occupied, empty or doubly occupied. This is not true however, since Eq.~\eqn{Gut-self-consistency-eq} connects the single-particle reduced density matrix of the bath levels with the 
local single-particle density matrix of the quasiparticles, which, in turn, affects the expectation value of $\mathcal{H}_*$ in Eq.~\eqn{gutenergy}. One can readily convince oneself that the lowest 
variational energy is obtained with the decoupled bath level singly-occupied and coupled 
in a spin-singlet configuration with the impurity+LHU+UHB or, in other words, the  
wavefunction 
\beal
\ket{\phi} &= \fract{1}{\sqrt{2}}\,\big( \ket{\psi_\up}\ket{\Psi_\down}
- \ket{\psi_\down}\ket{\Psi_\up}\big)\,. 
\label{impurity wavefunction}
\eal
Indeed, let us denote the Fock states of a single level 
as $\gamma=0$, the empty state, $\gamma=\sigma$, the state with a spin-$\sigma$ 
electron, and $\gamma=2$, the doubly occupied one, and the Fock states 
of the impurity+LHB+UHB as $\ket{\gamma_{imp},\gamma_\text{LHB},\gamma_{UHB}}$. 
For large $U$, we can safely assume the trial wavefunction   
\be
\ket{\Psi_\sigma} = \cos\theta\ket{\sigma,2,0} 
+\fract{\sin\theta}{\sqrt{2}}\,\big(
\ket{0,2,\sigma}+\ket{2,\sigma,0}\big)\,,\label{trial-wf}
\ee
with $\theta\simeq 0$. 
Using the spin-singlet configuration in Eq.~\eqn{impurity wavefunction}, we 
find that the occupation numbers of LHB and UHB are spin-independent are read
\beal
n_{ \text{LHB} \sigma} &= 1-\fract{1}{4}\,\sin^2\theta\,,&
n_{ \text{UHB} \sigma} &= \fract{1}{4}\,\sin^2\theta\,,
\label{n-LHU-UHB}
\eal
while the impurity is half-filled. 
We note that Eq.~\eqn{Gut-self-consistency-eq} entails that the quasiparticle local density matrix is the p-h transform of the impurity bath one, thus that 
the impurity LHB and UHB transform, respectively, into the quasiparticle 
UHB and LHB. Since, for $a=\text{LHB},\text{UHB}$,
\beal
\bra{\phi} d^\dagger_{a \sigma }\,c^\dagga_\sigma\ket{\phi}
&= \fract{\sin\theta\,\cos\theta}{2\sqrt{2}}\;,
\label{hybridisation expectation}
\eal
and $n_{a \sigma }(1-n_{a \sigma }) = \sin^2\theta/4$ are independent of $a$, the wavefunction renormalisation factors are 
as well, and read~\cite{Nicola-ghost} 
\bealn
R_{a\sigma} &= \fract{\bra{\phi} d^\dagger_{a \sigma}\,c^\dagga_\sigma\ket{\phi}}
{\sqrt{n_{a \sigma }\,\big(1-n_{a\sigma }\big)}}\\
&= \fract{2}{\sin\theta}\;\fract{\sin\theta\,\cos\theta}{2\sqrt{2}}\simeq
\fract{1}{\sqrt{2}}\,.
\eal
The quasiparticle Hamiltonian Eq.~\eqn{qpham} is therefore
\beal
\mathcal{H}^{qp} &= \fract{1}{2}\,\sum_{\bk\sigma}\,\sum_{a,b=
\text{LHB},\text{HUB}} \ep(\bk)\,d^\dagger_{\bk a\sigma}\,d^\dagga_{\bk b\sigma}\\
&\qquad -\lambda\,\sum_{\bk}\,
\big(n_{\bk\text{LHB}}-n_{\bk\text{UHB}}\big)\\
&= \mathcal{H}_* -\lambda\,\sum_{\bk}\,
\big(n_{\bk\text{LHB}}-n_{\bk\text{UHB}}\big)
\,,
\label{Hqp-simple}
\eal
plus the flat zero-energy band that, because of Eq.~\eqn{Gut-self-consistency-eq} 
and Eq.~\eqn{impurity wavefunction}, is half-filled with equal number of spin up and down quasiparticles. The Hamiltonian in Eq.~\eqn{Hqp-simple} describes two bands with dispersion 
\bealn
\ep_\mp(\bk) &= \fract{1}{2}\,\bigg(\ep(\bk)
\mp \sqrt{\ep(\bk)^2 + 4\lambda^2\;}\,\bigg)\,,
\eal
and corresponding eigenoperators
\beal
d^\dagga_{\bk\, -\,\sigma} &= \cos\theta_\bk\, 
d^\dagga_{\bk\text{LHB}\sigma} + \sin\theta_\bk\, 
d^\dagga_{\bk\text{UHB}\sigma}\,,\\
d^\dagga_{\bk\, +\,\sigma} &= -\sin\theta_\bk\, 
d^\dagga_{\bk\text{LHB}\sigma} + \cos\theta_\bk\, 
d^\dagga_{\bk\text{UHB}\sigma}\,,
\label{eigenoperators}
\eal
with $\tan 2\theta_\bk = \ep(\bk)/2\lambda$. 
The parameter $\lambda$ in Eq.~\eqn{Hqp-simple} is a Lagrange multiplier that enforces the self-consistency 
condition in Eq.~\eqn{Gut-self-consistency-eq}, namely, that the LHB is almost full and the 
HUB almost empty. This requires $\lambda\gg t$, thus $\ep_-(\bk)<0$ and $\ep_+(\bk)>0$ separated 
by a gap $2\lambda$, and the ground state $\ket{\Psi_*}$ obtained by filling the lowest-energy band. 
Specifically, the self-consistency Eq.~\eqn{Gut-self-consistency-eq} 
together with Eq.~\eqn{n-LHU-UHB} imply, through Eq.~\eqn{eigenoperators}, that  
\bealn
\fract{1}{V}\sum_\bk\,\bra{\Psi_*} n_{\bk\text{LHB}}\ket{\Psi_*}
&\simeq 2 -\fract{3t^2}{4\lambda^2}= 2-\fract{\sin^2\theta}{2}\,,\\
\fract{1}{V}\sum_\bk\,\bra{\Psi_*} n_{\bk\text{UHB}}\ket{\Psi_*}
&\simeq \fract{3t^2}{4\lambda^2}= \fract{\sin^2\theta}{2}\,.
\eal
It follows that $3t^2/2\lambda^2=\sin^2\theta$ and that the variational energy 
per site is 
\bealn
E &= \fract{1}{V}\,\bra{\Psi_*}\mathcal{H}_*\ket{\Psi_*} +  
\fract{U}{2}\,\bra{\phi}(n-1)^2\ket{\phi}\\
&\simeq -\fract{3t^2}{\lambda} +\fract{U}{2}\,\sin^2\theta 
= -\fract{3t^2}{\lambda}+\fract{U}{2}\;\fract{3t^2}{2\lambda^2}\,,
\eal
with minimum at $\lambda=U/2$ and value $E=-3t^2/U$. We recall that the 
half-filled Hubbard model at very large $U$ maps onto a spin-1/2 Heisenberg 
model
\beal
H_\text{Heis} &= J\,\sum_{<ij>}\,\bigg(
\bd{S}_i\cdot\bd{S}_j - \fract{1}{4}\bigg)\,,
\label{Heisenberg}
\eal
where $J=4t^2/U$. We realise that on a cubic lattice the large-$U$ gGut variational energy per site, $E=-3t^2/U$, actually $E=-zt^2/2U$ for generic lattices with coordination number $z$, reproduces the second term in parenthesis. This represents a big achievement with respect to the standard Gut, which yields a vanishing variational energy~\cite{Brinkman&Rice-PRB1970}, and is consistent with the fact that 
gGut is strictly variational for lattices with infinite coordination, where 
the term $\bd{S}_i\cdot\bd{S}_j$ does not contribute unless breaking the 
spin $SU(2)$.\\   
If we instead use as starting point of the optimization process  
the spin-triplet state, e.g., 
$\ket{\psi_\up}\ket{\Psi_\up}$, rather than the spin-singlet one in Eq.~\eqn{impurity wavefunction}, or the configurations in which the decoupled bath level is empty or doubly occupied,  
and repeat the previous calculations, we find at large $U$ the same solution as in standard Gut.
This, as mentioned above, results in a higher variational energy. The same occurs if the number of auxiliary orbitals, i.e., of bath levels, is $N_{aux}=2$. More generally, the variational solution in the Mott insulator with even $N_{aux}=2M$ coincides with that of $N_{aux}=2M-1$. In other words, the self-consistency condition Eq.~\eqn{Gut-self-consistency-eq} effectively provides a strong spin entanglement~\cite{guerci2019} between the impurity+LHU+UHB and the decoupled bath level, which is forced to lie at zero energy, thus yielding the zero-energy quasiparticle flat-band that, we argue, hosts the spinons.

\subsection{Ferromagnetic solution}
To support this conjecture, let us study a gGut variational wavefunction with 
finite magnetisation $m$ per site. Instead of Eq.~\eqn{impurity wavefunction}, see Eq.~\eqn{trial-wf}, we use as trial wavefunction of the impurity model 
\beal
\ket{\phi} &= \cos\theta\,\cos\phi \ket{\psi_\down}\ket{\up,2,0}\\
&\qquad - \cos\theta\,\sin\phi \ket{\psi_\up}\ket{\down,2,0}\\
&\qquad + \fract{\sin\theta}{2} \ket{\psi_\down}
\big(\ket{0,2\up}+\ket{2,\up,0}\big)\,,\\
&\qquad - \fract{\sin\theta}{2}\ket{\psi_\up}\big(\ket{0,2\down}+\ket{2,\down,0}\big)\,,
\label{impurity-wf-m}
\eal
which allows for a finite impurity magnetisation  
that corresponds to the physical one $m$, thus 
${n_\up-n_\down = \cos 2\phi\,\cos^2\theta = m}$.
We note that the magnetisation of the decoupled 
bath level coincides with $-m$, which shows that the physical spin is 
entirely carried by the zero-energy states in the quasiparticle Hamiltonian, whose magnetisation is the p-h transform of the decoupled bath level, thus precisely $m$. 
Indeed, the choice in Eq.~\eqn{impurity-wf-m} is intentionally done with that 
purpose, We have confirmed that it reproduces the numerically optimized 
solution. Correspondingly, the occupation numbers of the LHB and UHB are
the same as in Eq.~\eqn{n-LHU-UHB}, i.e., spin independent. 
What changes are the expectation values in Eq.~\eqn{hybridisation expectation}, 
which now read
\beal
\bra{\phi} d^\dagger_{\text{UHB} \up }\,c^\dagga_\up\ket{\phi}
&= \bra{\phi} d^\dagger_{\text{LHB} \down }\,c^\dagga_\down\ket{\phi}\\
&= \cos\phi\;\fract{\sin\theta\cos\theta}{2}\;,\\
\bra{\phi} d^\dagger_{\text{LHB} \up }\,c^\dagga_\up\ket{\phi}
&= \bra{\phi} d^\dagger_{\text{UHB} \down }\,c^\dagga_\down\ket{\phi}\\
&= \sin\phi\;\fract{\sin\theta\cos\theta}{2}\;,
\eal
resulting in the following wavefunction renormalisation factors
\beal
R_{\text{UHB} \up } &= R_{\text{LHB} \down }\simeq \cos\phi\,,\\
R_{\text{LHB} \up } &= R_{\text{UHB} \down }\simeq \sin\phi\,.
\label{R-m}
\eal
The consequence, as one can readily derive following the former calculations at $m=0$, is a reduction of 
\bealn
\bra{\Psi_*}\mathcal{H}_*\ket{\Psi_*} \simeq -\fract{3t^2}{\lambda}\,\sin^2 2\phi
\,,
\eal
and a modified self-consistency condition 
\bealn
\fract{3t^2}{2\lambda^2} &= \fract{\sin^2\theta}{\sin^2 2\phi}\;,
\eal
both of which reduce to the previous ones when $\phi=\pi/4$.
The variational energy per site is therefore
\bealn
E &= -\fract{3t^2}{\lambda}\,\sin^2 2\phi + \fract{U}{2}\,\sin^2 \theta\\
&= -3t^2\,\sin^2 2\phi\,\bigg(\fract{1}{\lambda} 
- \fract{U}{4\lambda^2}\bigg) 
\,,
\eal
with the constraint 

\bealn
m &= \cos 2\phi\,\cos^2\theta = \cos 2\phi\,
\bigg(1-\fract{3t^2}{2\lambda^2}\,\sin^2 2\phi\bigg)\\
&\xrightarrow[\lambda\gg t]{} \cos 2\phi
\,,
\eal

from which we find that the optimal $\lambda=U/2$, as before, and 
\beal
E(m) &= -\fract{3t^2}{U}\,\big(1-m^2\big)\,.
\label{E(m)}
\eal
We note that Eq.~\eqn{E(m)} corresponds precisely to the expectation value per site 
of the Heisenberg Hamiltonian in Eq.~\eqn{Heisenberg} over a state 
with average spin polarization $\langle S^z_i\rangle = m/2$
in the limit of infinite coordination-number. In other words, in this limit the variational gGut wavefunction reproduces the exact ground state energy and it does so in a very physical way: the spin polarization is carried just by the quasiparticles of the flat band, justifying our interpretation of them as 
\textit{spinons}, whereas the dispersive Hubbard bands remain unpolarized. \\
We further observe that, even though the spinon band 
is flat in the quasiparticle Hamiltonian, the variational energy in Eq.~\eqn{E(m)} is instead compatible with a non-zero dispersion and a finite spinon density-of-states at zero energy. 
The reason lies in the fact that 
the spinon and Hubbard bands are not truly independent of each other. Indeed, the spinon polarization does affect the Hubbard bands through 
\eqn{R-m}, and, in turn, the expectation value $\bra{\Psi_*}\mathcal{H}_*\ket{\Psi_*}$. This behaviour is similar to the outcome of conventional Hartree-Fock plus random-phase approximation, and suggests that the inclusion of quantum fluctuations, as done for the Gut using the time-dependent variational principle~\cite{fluctuations}, may explicitly unveil the coupling between 
the Hubbard bands and the spinon one. In that case, a finite spinon dispersion should 
appear upon integrating out the high-energy Hubbard bands. Moreover, the finite spinon density-of-states consistent with Eq.~\eqn{E(m)} suggests that the spinons do have a well-defined Fermi-surface.

\subsection{Antiferromagnetic solution}  
To conclude this section and provide further evidence of the quality of 
the variational gGut wavefunction, let us briefly discuss the solution when we allow for antiferromagnetism. On a bipartite lattice with hopping just between the two sublattices, we expect the ground state to describe an antiferromagnetic insulator for any $U>0$, as mentioned in the Introduction. Let us therefore extend the previous analysis to an  antiferromagnet at fixed staggered magnetisation $m$. We assume that both sublattices, which we denote as A and B, are 
described by the auxiliary impurity wavefunction \eqn{impurity-wf-m} with 
sublattice dependent $\phi$, specifically, $\phi_A=\phi$ and $\phi_B=\pi/2-\phi$, 
thus opposite sublattice magnetisations
\beal
m_A &= \cos2\phi\,\cos^2\theta \equiv m\,,&
m_B&=-m\,.
\label{AF-m}
\eal 
The occupation numbers per spin are still those in \eqn{n-LHU-UHB}. The wavefunction renormalisations are the same as in \eqn{R-m} for sublattice A, while 
$\cos\phi\leftrightarrow\sin\phi$ for sublattice $B$. It follows that the quasiparticle  Hamiltonian \eqn{qpham} in the LHB-UHB orbital space is characterised by uniform diagonal and off-diagonal hopping amplitudes $-t\,\sin2\phi/2$ and $-t/2$, respectively, as well as by a staggered off-diagonal hopping $-t\,\cos2\phi/2$ due to antiferromagnetism. In addition, the Hamiltonian includes a Lagrange multiplier $\lambda$ that enforces 
the self-consistency condition $n_\text{LHB}-n_\text{UHB}=2-\sin^2\theta$, and one that fixes the local density, $n_\text{LHB}+n_\text{UHB}=2$. For $U\gg t$, $\lambda$ is also large and the quasiparticle Hamiltonian $\mathcal{H}^{qp}$ in the halved, magnetic Brillouin zone (MBZ) describes 
an insulator with occupied valence bands
\bealn
\epsilon_1(\bk) &= -\frac{1}{2} \sqrt{2\epsilon(\bk)^2 + 4\lambda^2 -2\epsilon(\bk)\,E(\bk)\;} \\
\epsilon_2(\bk) &= -\frac{1}{2} \sqrt{2\epsilon(\bk)^2 + 4\lambda^2 +2\epsilon(\bk)\,E(\bk)\;} \,,  
\eal
and empty conductions ones, $\ep_3(\bk)=-\ep_2(\bk)$ and $\ep_4(\bk)=-\ep_1(\bk)$,  
where $E(\bk) =\sqrt{\epsilon(\bk)^2 + 4\lambda^2 \sin^2 2\phi\;}$.
The constraint $n_\text{LHB}+n_\text{UHB}=2$ is therefore automatically satisfied. 
Concerning the other constraint, if 
\bealn
E^{qp}(\lambda) &= \fract{2}{V}\,\sum_{\bk\in\text{MBZ}}\,\big(\ep_1(\bk)+\ep_2(\bk)\big)\\
&\simeq -2\lambda -\big(1+\cos^2 2\phi\big)\,
\fract{3t^2}{2\lambda}
\,,
\eal
is the ground state energy per site of $\mathcal{H}^{qp}$, the last equation being valid for $\lambda\gg t$ on a cubic lattice, we have to impose that 
\bealn
n_\text{LHB}-n_\text{UHB}&=2-\sin^2\theta = -\fract{\partial E^{qp}(\lambda)}{\partial\lambda}\\
&\simeq 2 -\big(1+\cos^2 2\phi\big)\,
\fract{3t^2}{2\lambda^2}\;,
\eal
thus
\bealn
\sin^2\theta &= \big(1+\cos^2 2\phi\big)\,
\fract{3t^2}{2\lambda^2}\;,
\eal
and 
\bealn
\fract{1}{V}\,\bra{\Psi_*}\mathcal{H}_*\ket{\Psi_*} &= 
E^{qp}(\lambda) + \lambda\,\big(n_\text{LHB}-n_\text{UHB}\big)\\
&\simeq -\fract{3t^2}{\lambda}\,\big(1+\cos^2 2 \phi\big)\,.
\eal
Performing the same steps as in the calculations in the ferromagnetic case, we finally get a variational energy 
per site that reads, in the general case of a $d$-dimensional hypercubic lattice,  
\beal
E_\text{AF}(m) &= -\fract{dt^2}{U}\,\big(1+m^2\big)\,,
\label{E(m) AF}
\eal
which decreases with $|m|$, showing that the paramagnetic solution at $m=0$ is unstable towards developing an antiferromagnetic order parameter. Moreover, \eqn{E(m) AF} coincides with the ground state energy of \eqn{Heisenberg} at fixed staggered magnetisation $m$ for large $d$, i.e., when quantum fluctuations can be ignored. 
As before, the staggered magnetisation is carried solely by the spinons, the flat-band 
quasiparticles, while the Hubbard bands are spin unpolarised. This evidently provides a much better description of the large $U$ antiferromagnetic insulator than, e.g., the Hartree-Fock approximation, where the staggered magnetisation is due to the valence and 
conduction bands that mimic, within Hartree-Fock approximation, the Hubbard bands, and are separated from each other by a gap $\sim 2U m$ instead of the expected $U$, the actual cost of a charge excitation.

\section{Conclusions}
 \label{Conclusions}
 We have shown that the ghost Gutzwiller approximation is able to 
 describe a genuine paramagnetic Mott insulator that hosts neutral spin-1/2 quasiparticles,   
 possibly with their own Fermi surface, the sought-after 
 Anderson's spinons. 
Specifically, the spin-singlet wavefunction of the auxiliary impurity 
model implies that the linear operator $\mathcal{P}_G$ in \eqn{gutansatz} 
is spin $SU(2)$ invariant.
The spin state of the full ghost Gutzwiller wave function thus depends on
the spin state of the Slater determinant in Eq.~\eqref{gutansatz}, which our results 
suggest to be a singlet.
Indeed, the contribution of the Hubbard bands to the Slater determinant is trivially 
a spin-singlet state, since they are either completely full or empty.
While the energy optimization does not completely determine the spinon contribution to the Slater determinant,
as it appears with its corresponding $R$ equal to zero, we know that this term is characterized by an 
$SU(2)$-invariant local density matrix, fixed by the ghost Gutzwiller self-consistency.
We believe that these 
observations, together with the finite paramagnetic susceptibility entailed by \eqn{E(m)}, collectively provide evidence that the optimal ghost-Gutzwiller wavefunction in the Mott phase is indeed a global spin singlet with gapless spinons. \\   
Furthermore, we have demonstrated that these spinons can be revealed by a proximity effect at the interface between the paramagnetic Mott insulator and a metal. Indeed, 
 the spinons near the interface regain the electron charge and get promoted into  
 bona fide quasiparticles with a narrow dispersion in the directions parallel to the interface 
 that decreases exponentially moving inside the bulk of the Mott insulator.  
 This behavior resembles more a heavy-fermion proximity effect rather than a Kondo-like one that was advocated to interpret the DMFT results~\cite{Rosh&Costi-PRL2008}.\\
 Lastly, we have shown that when the system is allowed to have a finite magnetization $m$, the quadratic energy raise with $m$ is compatible with the spinons having a finite density of state hence a finite dispersion in the Brillouin zone, not accessible at $m=0$.\\
To unveil the precise form of the spinon dispersion one needs to include quantum fluctuations on top of the variational solution, a work that is currently underway.

\section*{Acknowledgments}
A.M.T. acknowledges useful discussions with Gregorio Stafferi.
\appendix
\section{Gutzwiller variational equations}\label{gutappendix}
In this section, we show the gGut equations in the case of a metal-Mott insulator interface. We hereafter impose just the spin-$U(1)$ symmetry corresponding to rotations around the spin $z$axis.\\
To optimize the wavefunction in~\ref{gutansatz} we use the Gutzwiller approximation, namely, we use 
results in the infinite coordination limit and apply them also to finite coordination lattices. 
We can analytically calculate the expressions of all expectation values over the variational 
Gutzwiller wavefunction, if we further impose the so-called Gutzwiller constraints~\cite{Bunemann1998,Fabrizio2007,lanata2015,fluctuations}, which, in the impurity model formulation,  read: 
\beal
\braket{\phi(\ell)|\phi(\ell)} &= 1\,,\\
\bra{\phi(\ell)} d^\dagga_{\ell\alpha\sigma}\,d^\dagger_{\ell\alpha\sigma}\ket{\phi(\ell)} &= \bra{\Psi_*}d^\dagger_{\ell\bm{r}\beta\sigma}\,  d^\dagga_{\ell\bm{r}\alpha\sigma} \ket{\Psi_*}\\
& \equiv \Delta_{\alpha\sigma,\beta\sigma}(\ell)\,,\label{App:constraints}
\eal
where $\Delta_{\alpha\sigma,\beta\sigma}(\ell)$ define a matrix $\hat{\Delta}_\sigma(\ell)$. 
More specifically, the minimization of the energy~\ref{gutenergy}, after imposing Eq.~\eqn{App:constraints} and within the Gutzwiller approximation can be done algebraically by solving the following system of equations:
\begin{widetext}
\beal
\mathcal{H}_{qp} \ket{\psi_*} &= E_{qp}\ket{\psi_*}~,\\
\bigg(\sqrt{\hat{\Delta}_\sigma(\ell)\big(1 -\hat{\Delta}(\ell)_\sigma\big)\,}\, 
    \bm{V}_\sigma(\ell)\bigg)_{\alpha} &= \sum_{\bk,\ell',\beta} R_{\beta\sigma}(\ell')\,\bra{\Psi_*} 
    d^\dagger_{\ell\bk\alpha\sigma}\,d_{\ell'\bk\beta\sigma}  \ket{\Psi_*}\, t_{\ell\ell'}(\bk)\,,\\
    \hat{\lambda}_{\sigma}(\ell) &= -\hat{\lambda}^{c}_{\sigma}(\ell) + \frac{\partial}{\partial \hat{\Delta}_{\sigma}(\ell)} \,\bigg(
    \bm{R}_\sigma(\ell)\cdot \sqrt{\hat{\Delta}_\sigma(\ell)\big(1 -\hat{\Delta}(\ell)_\sigma\big)\,}\, 
    \bm{V}_\sigma(\ell) + H.c. \bigg) \,,\\
\mathcal{H}_{imp}(\ell) \ket{\phi(\ell)} &= E_{imp}(\ell)\ket{\phi(\ell)}\,,\\
\bra{\phi(\ell)} d^\dagga_{\ell\alpha\sigma}\,d^\dagger_{\ell\alpha\sigma}\ket{\phi(\ell)} &= \bra{\Psi_*}d^\dagger_{\ell\bm{r}\beta\sigma}\,  d^\dagga_{\ell\bm{r}\alpha\sigma} \ket{\Psi_*}= \Delta_{\alpha\sigma,\beta\sigma}(\ell)\\
R_{\alpha\sigma}(\ell) &= \sum_{\beta}\bigg(\sqrt{\hat{\Delta}_\sigma(\ell)\big(1 -\hat{\Delta}(\ell)_\sigma\big)\,}\,\bigg)^{-1}_{\alpha,\beta}\, \bra{\phi(\ell)} c^{\dagger}_{\ell\sigma}\,
d^\dagga_{\ell\beta\sigma}\ket{\phi(\ell)}\,,\label{R}
\eal
\end{widetext}
where 
\bealn
t_{\ell\ell'}(\bk) &= \delta_{\ell\ell'}\,\ep(\bk) -t\,\big(\delta_{\ell',\ell+1}
+\delta_{\ell',\ell-1}\big)\,,
\eal 
$\bm{R}_\sigma(\ell)$ and $\bm{V}_\sigma(\ell)$ are vectors with 
components $R_{\alpha\sigma}(\ell)$ and $V_{\alpha\sigma}(\ell)$, respectively, while 
$\hat{\lambda}_{\sigma}(\ell)$ and $\hat{\lambda}^c_{\sigma}(\ell)$ matrices with 
elements $\lambda_{\alpha\sigma,\beta\sigma}(\ell)$ and 
$\lambda^c_{\alpha\sigma,\beta\sigma}(\ell)$, respectively. The solution of Eq.~\eqn{R} 
can be obtained self-consistently starting from a guess for $\bm{R}_\sigma(\ell)$ and $\hat{\lambda}_{\sigma}(\ell)$ and iterating the procedure 
till every equation is satisfied up to the desired accuracy.

\section{continuum description for the wetting critical behaviour}\label{continuum}
%\begin{widetext}
In the case of the Gut the Mott transition is first order but we still have a surface critical behaviour, as shown in the main text. In this case, we can find a continuum description to extract the critical behaviour, as done in~\cite{Borghi-PRB2010} at p-h symmetry. 
We write the impurity wavefunction as 
\bealn
\ket{\phi(\ell)} &= \phi_0(\ell)\,\ket{0,2} + \phi_2(\ell)\,\ket{2,0}\\
&\qquad + \sum_\sigma\, \sigma\,\phi_\sigma(\ell) \ket{\sigma,-\sigma}
\,,
\eal
where $\ket{\gamma_{imp},\gamma_{bath}}$ is a Fock state of the impurity plus the single bath level. 
Hereafter, we assume spin-$SU(2)$ symmetry, thus $\phi_\sigma(\ell) = \phi_1(\ell)$ independent 
of spin.  
For the standard Gutzwiller ansatz, we can write the energy functional per site as:
\bealn
E &= \fract{1}{N}\,\sum_\ell\,\bigg\{\frac{U_\ell}{2} \,\big(\phi_0(\ell)^2 + \phi_2(\ell)^2\big) -\delta\mu\, \delta(\ell)\bigg\}\\
& \qquad + \fract{1}{N}\sum_{\ell}\,R^2_\sigma(\ell)\,\ep_\ell \\
&\qquad + \fract{1}{N}\sum_{\ell}\,\sum_{p=\pm 1}\,\,R_\sigma(\ell)\,t_{\ell\ell+p}\,R_\sigma(\ell+p)\,,
\eal
where $N$ is the number of layers and 
\beal
\ep_\ell &= \fract{1}{\mathcal{A}}\,\sum_{\bk\sigma}\,\bra{\Psi_*} 
c^\dagger_{\ell\bk\sigma}\,\ep(\bk)\, c^\dagga_{\ell\bk\sigma}\ket{\Psi_*}\,,\\
t_{\ell\ell'} &= -\fract{t}{\mathcal{A}}\,\sum_{\bk\sigma}\,\bra{\Psi_*} 
c^\dagger_{\ell\bk\sigma}\,c^\dagga_{\ell'\bk\sigma}\ket{\Psi_*} 
\,.
\label{App:wetting}
\eal
The Gutzwiller constraints and the equation for $R_\sigma(\ell)=R(\ell)$ read:
\bealn
1&= \phi_0(\ell)^2 + \phi_2(\ell)^2 + 2\, \phi_1(\ell)^2 \,,\\
n(\ell) &= 1-\delta(\ell) = 2\phi_2(\ell)^2 + 2\phi_1(\ell)^2\,,\\
R(\ell)^2 &= \fract{4}{1-\delta(\ell)^2} \big(\phi_2(\ell)\,\phi_1(\ell) + \phi_0(\ell)\,\phi_1(\ell)\big)^2\,.
\eal
We can choose to express the energy functional using $R(\ell)$ and the doping $\delta(\ell)$ as independent variables, in particular, 
\bw
\beal
\phi_0(\ell)^2 + \phi_2(\ell)^2 &= \frac{1}{2}\Bigg(1 - \sqrt{\big(1 - R(\ell)^2\big)\big(1 - \delta(\ell)^2\big)\,} + \frac{\delta(\ell)^2}{\;1-\sqrt{\big(1 - R(\ell)^2\big)\big(1 - \delta(\ell)^2\big)\,}\;}\Bigg)\\
 &\simeq \frac{1}{2}\Bigg(1 - \sqrt{1 - R(\ell)^2\,} +\frac{\delta(\ell)^2}{2}\;
 \frac{\;1+ R(\ell)^2 + \sqrt{1 -R(\ell)^2\,}\;}{\;1-\sqrt{1 - R(\ell)^2\,}\;}\,\Bigg)\,.
 \label{App:intermediate}
\eal
\ew
Where the last equation is valid in the limit of small doping.\\
Near the critical point, the quantities in Eq.~\eqn{App:wetting} do not depend appreciably 
on $\ell$, $\ep_\ell \simeq -\ep_\parallel$ and $t_{\ell\ell\pm 1}\simeq -\ep_\perp/2$, so that, in the continuum limit $\ell\to x$, with continuous $x$, the energy functional reads 
\beal
  E &= \int {\it dx} \Bigg\{\frac{U(x)}{2}\,\Big[\phi_0(x)^2+\phi_2(x)^2\Big] - \epsilon\, R(x)^2 \\
  &\qquad\qquad\qquad\quad\;\; + \frac{\ep_\perp}{2}\, \left(\frac{\partial R(x)}{\partial x}\right)^2 -\delta\mu\, \delta(x) \Bigg\}\,,
\eal
where $\ep=\ep_\parallel +\ep_\perp$, and $\phi_0(x)^2+\phi_2(x)^2$ is given by Eq.~\eqn{App:intermediate}.
Solving the Euler-Lagrange equations we can express $\delta(x)$ as a function of R(x):
\beal
    \delta(x) = \frac{4\delta\mu}{U(x)}\frac{1 - \sqrt{1-R(x)^2}}{1 + R(x)^2 + \sqrt{1-R(x)^2}}~.
\label{delta_x}
\eal
The solution for $R(x)$ can be found deep in the Mott insulator, where $U(x)=U$ is constant and 
$R(x)$ and $\delta(x)$ are 
small. We find: 
\bealn
    R(x) &\simeq \esp{-x/\xi}\;,&
    \delta(x) &\simeq \text{sign}(\delta\mu)\,R(x)^2
\eal
with the correlation length given by
\bealn
\fract{1}{\xi^2} &= \fract{1}{\ep_\perp}\,\bigg(\fract{U}{4} -2\ep -\fract{\delta\mu^2}{U}\bigg)\,.
\eal
We note that $\xi$ diverges at a critical $U=U_c$ which shifts to higher values at $\delta\mu\not=0$, 
as expected.  

\end{document}